%% file: spsc_v2-globalsip.tex
\documentclass[conference, 10pt]{IEEEtran}
\usepackage{import}
\usepackage{footnote}
\usepackage{flushend}
\usepackage{pbox}
\usepackage{amsmath}
\usepackage{amssymb}
\usepackage{dcolumn}
\usepackage{hyperref}
\usepackage{pgfplots}
\definecolor{deepgreen}{rgb}{0.1,0.6,0.1}
\usetikzlibrary{plotmarks}
\hypersetup{
pdfauthor={Alexandre J. Raymond and Warren J. Gross},%
pdftitle={Scalable Successive-Cancellation Hardware Decoder for Polar Codes},
colorlinks=true,
linkcolor=blue,
citecolor=deepgreen
}%
\graphicspath{ {./fig/} }

\ifCLASSINFOpdf
\else
\fi
\usepackage{array}

\usepackage{stfloats}
\usepackage{graphicx}
\usepackage{color}
\usepackage{placeins}
\usepackage{tabularx,colortbl}

\input{defines}

\begin{document}
\title{Scalable Successive-Cancellation Hardware\\Decoder for Polar Codes}

\author{\IEEEauthorblockN{Alexandre J. Raymond and Warren J. Gross}
\IEEEauthorblockA{Department of Electrical and Computer Engineering\\
McGill University\\
Montr\'eal, Qu\'ebec, Canada\\
alexandre.raymond@mail.mcgill.ca, warren.gross@mcgill.ca\\}}


%Telephone: (800) 555--1212, Fax: (888) 555--1212}


\maketitle

\begin{abstract}
%\boldmath
Polar codes, discovered by \Arikan{}, are the first error-correcting codes with an explicit construction to provably achieve channel capacity, asymptotically. However, their error-correction performance at finite lengths tends to be lower than existing capacity-approaching schemes. Using the successive-cancellation algorithm, polar decoders can be designed for very long codes, with low hardware complexity, leveraging the regular structure of such codes. We present an architecture and an implementation of a scalable hardware decoder based on this algorithm. This design is shown to scale to code lengths of up to $N=2^{20}$ on an Altera Stratix~IV FPGA, limited almost exclusively by the amount of available SRAM.
\end{abstract}


% \fi
\IEEEpeerreviewmaketitle

\section{Introduction} 
Since their introduction in 2008, polar codes \cite{Arikan2008} have attracted a lot of attention from the information theory community, as they are the first codes to provably achieve channel capacity, asymptotically in code length.

Although initially only defined for the binary erasure channel (BEC), they were later extended to other models, such as the additive white Gaussian noise (AWGN) channel \cite{Tal2011b}.

Their recursive construction was shown to support low-complexity implementations of the successive-cancellation (SC) algorithm in hardware \cite{Leroux_TSP_2012}\cite{Mishra_ASSCC_2012}. Those low-complexity decoders can in turn be used as components in more complex schemes, such as list decoding \cite{Tal2011}\cite{Balatsoukas2013} and concatenated coding \cite{Samsung2013}, which improve the error-correction performance of polar codes at finite lengths.

The remainder of this paper is structured as follows. \secref{sec:background} provides background information on polar codes and SC decoding. Then, \secref{sec:architecture} details the proposed architecture. \secref{sec:experimental_results} analyzes FPGA implementation results, while \secref{sec:conclusion} concludes this work.

\input{fig-qtz_perf}

\section{Background}
\label{sec:background}
Polar codes are a class of linear block codes based on a recursive definition. They are constructed using a generator matrix $\mathbf{G}$, obtained from the base matrix $\mathbf{F_2}=\bigl(\begin{smallmatrix}1&0\\1&1\end{smallmatrix}\bigr)$, using
\begin{equation*}
\mathbf{G}
= \mathbf{F_N}
\triangleq \mathbf{(F_2)^{\otimes n}},
%= \begin{bmatrix}
%1 & 0 \\
%1 & 1 \\
%\end{bmatrix}^{\otimes n}
\end{equation*}
where $N=2^n$ is the code length, and $\otimes$ represents the Kronecker product. In this paper, we use $\bfu$ to denote an information vector, $\bfx$ for a codeword, $\bfy$ for a received vector, and $\bfuhat$ for the information vector estimated by the decoder.

These codes can be decoded using a recursive, multi-stage structure featuring $n$ stages of $N/2$ nodes, yielding a complexity $\mathcal{O}(N \log N)$ \cite{Arikan2008}.

To simplify their implementation in hardware, decoding can be carried out in the log-likelihood-ratio (LLR) domain, where the SC equations become the standard sum-product algorithm (SPA) equations, which can be approximated using the well-known min-sum algorithm (MSA) \cite{Leroux_ICASSP_2011}:
\begin{align}
&\LLR{f}(\LLR{a},\LLR{b})         \approx \Signr{\LLR{a}}  \Signr{\LLR{b}} \min (\Mag{\LLR{a}},\Mag{\LLR{b}}),
\label{eq:llr_f}\\
&\LLR{g}(\Hs{},\LLR{a},\LLR{b}) = \LLR{a}(-1)^{\Hs{}}+\LLR{b},
\label{eq:llr_g}
\end{align}
where $\Hs{}$ designates a partial sum. This approximation yields a performance degradation of $\sim$0.1dB over SPA, as illustrated in \figref{fig:qtz_perf}, although this gap tends to shrink for higher-rate ($R=k/N$) codes.

\begin{figure*}[t]
\centering
\noindent
\includegraphics[width=7in]{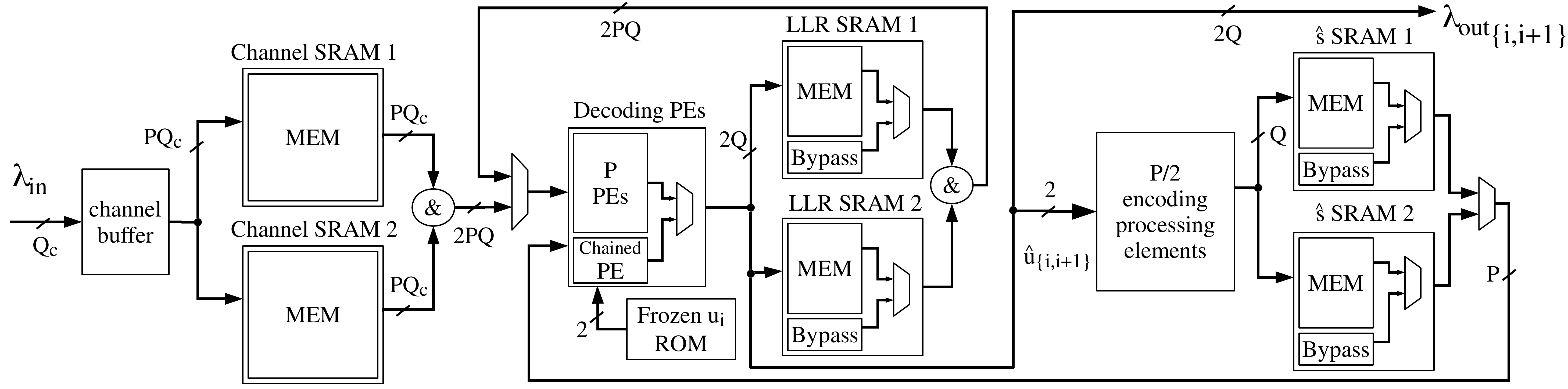}
\caption{Block diagram of the improved decoder architecture.}
\label{fig:toplevel2}
\end{figure*}

\section{Architecture}
\label{sec:architecture}
The architecture presented in this paper is based on the semi-parallel decoder of \cite{Leroux_TSP_2012}, and introduces modifications aiming to improve its scalability with respect to code length. This decoder uses a fixed datapath, and operates under resource constraints, where only $P \ll N/2$ processing elements (PE) are implemented. This limitation, however, only impacts throughput minimally \cite{Leroux_ICASSP_2011}.

\figref{fig:toplevel2} provides a top-level overview of the redesigned decoder architecture, while its various changes are discussed in the following sections.

\subsection{Memory Improvements}
\label{sec:mem_improvements}
Unlike \cite{Leroux_TSP_2012}, which makes use of a single SRAM to store all LLRs, this improved architecture relies on two separate types of memories: channel and internal. This separation allows full-throughput operation of the decoder by supporting the loading of a subsequent frame into the channel memory, without write contention, while the previous one is still being processed. This is made possible by the fact that, per the structure of the decoding graph, channel LLRs are not directly required in the second half of the decoding process, i.e. after bit $i=N/2$ and stage $l=(n-1)$.

Furthermore, the improved design does away with asymmetric read/write ports in its SRAMs. Those memories are replaced by pairs of $P$-LLR wide SRAMs, whose outputs are concatenated into $2P$-LLR words consumed by the processing elements, whose own $P$-LLR outputs are written to each SRAM in sequence. Note that the \texttt{\&} operator used in \figref{fig:toplevel2} symbolizes concatenation, with sign extension if needed.

\begin{figure}[t]
\centering
\noindent
\includegraphics[width=\columnwidth]{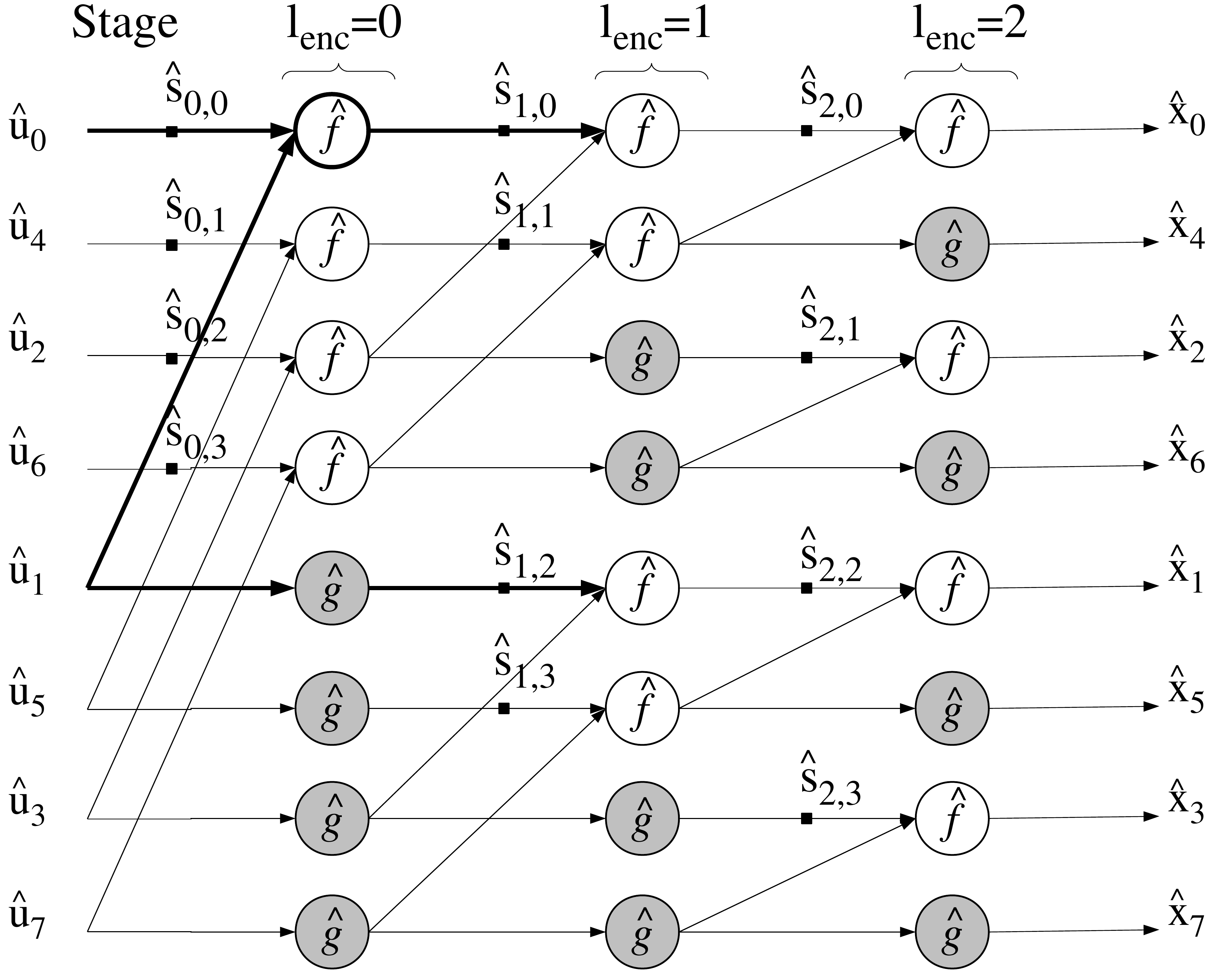}
\caption{Encoding graph for the partial sums.}
\label{fig:FFT_encoder_brev}
\end{figure}

\subsection{Quantization}
\label{sec:quantization}
The separation of the channel and internal memories, described in \secref{sec:mem_improvements}, also makes it possible to use distinct quantization levels for each memory. This enhancement is suggested by a characteristic of the successive-cancellation algorithm, namely that \eqref{eq:llr_g} affects the range of the computations in each successive stage, while their precision remains unchanged by both operations. It follows that the values processed by lower-indexed stages require more range than those in the higher ones.

Since the decoder must retain an entire $N$-LLR frame in memory, the channel SRAMs account for nearly half of the decoder's soft information storage requirements \cite{Leroux_TSP_2012}. A lower quantization for this memory therefore reduces the decoder area significantly.

Quantization is denoted using shorthand $(Q_i,Q_{i_c},Q_f)$, which indicates the number of integer bits for internal LLRs, integer bits for channel LLRs, and fractional bits for both types, respectively; $Q=Q_i+Q_f$ and $Q_c=Q_{i_c}+Q_f$ are also used to refer to the total number of quantization bits in each case.

Simulations showed that full-range quantization does not benefit error-correction performance; much lower levels can match a floating-point implementation. Specifically, we carried out those simulations for codes of length $N=2^{15}$, with $R \in \{0.25, 0.50, 0.75, 0.90\}$; results are summarized in \tblref{tbl:qtz_2_15}. We found that, for those codes, 6--8 bits of quantization suffice for good error-correction performance (within $\sim$0.1dB of floating point MSA), depending on their rate, as shown in \figref{fig:qtz_perf}. We also noticed that higher-rate codes tend to require fewer bits of fractional precision and integer range for internal LLRs, but more bits of integer range for channel ones.

\begin{table}[h]
\centering
\caption{Quantization required for good error-correction performance of $N=2^{15}$ codes using MSA.}
\label{tbl:qtz_2_15}
%\resizebox{3in}{!}{
\begin{tabular}{|c||c|c|c|c|}
\hline
$R$                   & 0.25      & 0.50      & 0.75      & 0.90\\
\hline
$(Q_i, Q_{i_c}, Q_f)$ & $(6,3,2)$ & $(6,3,2)$ & $(6,4,1)$ & $(6,4,0)$\\
\hline
\end{tabular}
%}
\end{table}

\subsection{Chained PE}
This architecture makes use of a chained PE in stage $0$, carrying out functions $\LLR{f}$ and $\LLR{g}$ in a single clock cycle (CC). The concept behind this improvement was introduced in \cite{Zhang2012}, while the restricted implementation used in this paper, targeting only stage $0$, was independently proposed in \cite{Mishra_ASSCC_2012}.

This chained PE relies on the specific schedule of the polar decoding graph, in which stage $0$ is always activated twice in a row, using the same operands: first for function $\LLR{f}$, and then for function $\LLR{g}$, using the result of $\LLR{f}$. By chaining both operations in a special PE, we can output two decoded bits $\Hu{\{i,i+1\}}$ at once, yielding a $(N/2)$-CC reduction in decoding latency.

This behavior is illustrated in \tblref{tbl:schedule}, specifically in clock cycles $\{3,6,13,16\}$. In those cases, the computations of functions $\LLR{f}$ and $\LLR{g}$ are performed in the same clock cycle, yielding two decoded bits simultaneously.

The chained PE does not incur any overhead over the regular PE. The data dependency present in-between functions $\LLR{f}$ and $\LLR{g}$, satisfied by the sign of $\LLR{f}$, occurs late in the processing of $\LLR{g}$, and can be computed very rapidly.

\subsection{Semi-Parallel Partial-Sum Encoder}
The main factor limiting the scalability of \cite{Leroux_TSP_2012} is the growing complexity of its partial-sum update logic. In this paper, we introduce an encoder-based alternative inspired by the design of \cite{Zhang2012}, which proposed a fully-parallel partial-sum computation module. Our implementation extends this encoder, adapting it to a novel semi-parallel architecture. This architecture operates over multiple clock cycles and uses a fixed datapath, removing it from the decoder's critical path altogether.

This encoder is triggered after decoding-stage $0$, and processes two decoded bits at a time. \figref{fig:FFT_encoder_brev} illustrates its structure, a mirrored version of the decoding graph, in which the $\Hf{}$ nodes are defined as binary additions (XOR), and the $\Hg{}$ nodes, as pass-through connections:
\begin{align}
\Hf{}(\Hu{a},\Hu{b}) &= \Hu{a} \oplus \Hu{b},\\
\Hg{}(\Hu{a})        &= \Hu{a}.
\end{align}
As in the decoding graph, the nodes are associated into $N/2$ pairs per stage. Those pairs are processed by the $P/2$ encoding PEs.

In order to make the design scalable, a semi-parallel architecture was chosen for the encoder. Since the encoding graph mirrors the decoding graph, their schedules are very similar. The encoding schedule is illustrated in \tblref{tbl:schedule}, where $e$ denotes the activation of encoding stage $l_\text{enc}$. Due to the semi-parallel nature of the encoder, stages which are handled in multiple clock cycles are denoted using a subscript, e.g. $e_0$.

In \figref{fig:FFT_encoder_brev}, the subgraph highlighted in bold illustrates the nodes activated to calculate partial sums $\Hs{1,0}$ and $\Hs{1,2}$. Those two values are subsequently used to evaluate $\LLR{g}$ nodes in stage $l=1$ of the decoding graph.

The partial-sum encoder follows a schedule similar to that of the decoding, although with half as many processing elements; those processing elements produce two values instead of one, since they are not restricted by a data dependency as the decoding PEs are. The encoder thus increases latency by $(\frac{N}{P}(P-1) + \frac{N}{P} \log_2(\frac{N}{4P}) - \log_2 P + 2)$ CC, or $\sim$67\% for $P=64$, but allows higher operating frequencies, for a net throughput gain.

Using $P/2$ encoding PEs, the encoder can make use of $P$-bit wide words in the $\Hs{}$ SRAMs, allowing the decoder to retrieve $P$ partial sums simultaneously during decoding, in a single clock cycle. Furthermore, because of the specific structure of the encoding graph, the values stored in memory are properly aligned for direct consumption by the decoding PEs, via a fixed datapath.

Note that the internal partial sums $\Hs{0,j}$ correspond to $\Hu{i}$, where $i$ is the bit-reversed \cite{Arikan2008} value of $j$. Furthermore, $\Hs{n,j}$ yields an estimation of codeword value $\Hx{i}$, where $i$ is again bit-reversed $j$. As part of the encoding process resulting in this estimated codeword $\bfxhat{}$, the encoder creates internal estimations $\Hs{l,j}$, which are required by $\LLR{g}$ during the decoding process.

In a non-systematic polar decoder, it is not necessary to evaluate $\bfxhat{}$ completely, which saves a final encoding stage after $\Hu{N-2}$ and $\Hu{N-1}$ are decoded. However, in a systematic decoder \cite{Arikan2011}, those extra steps could be carried out to obtain $\bfxhat$, which is required to retrieve the original information vector, while avoiding the need for extra hardware to perform the additional encoding step.

\begin{table}[t]
\setlength{\extrarowheight}{2.5pt}
\centering
\caption{FPGA implementation results targeting the Altera Stratix~IV GX EP4SGX530KH40C2.}
\label{tbl:fpga_results}
\resizebox{3.5in}{!}{
\begin{tabular}{|c|c|c|c|| D{;}{}{3}|D{;}{}{3}|D{;}{}{3}|c|c|}
\hline
$N$ & $R$ & $P$  & Qtz. & \multicolumn{1}{|c|}{LUT}    & \multicolumn{1}{|c|}{FF}     & \multicolumn{1}{|c|}{\pbox{1cm}{SRAM\\(bits)}} & \pbox{1cm}{$f_\text{max}$\\(MHz)} & \pbox{1cm}{T/P\\(Mbps)}\\
\hline
 $2^{15}$  & 0.25 & 64  & (6,3,2) &  4,;161  & 1,;629 &    510,;464 & 156 & $15$\\
 $2^{15}$  & 0.50 & 64  & (6,3,2) &  4,;161  & 1,;629 &    510,;464 & 156 & $29$\\
 $2^{15}$  & 0.75 & 64  & (6,4,1) &  3,;731  & 1,;496 &    477,;440 & 155 & $43$\\
 $2^{15}$  & 0.90 & 64  & (6,4,0) &  3,;263  & 1,;304 &    411,;648 & 167 & $56$\\
\hline
 $2^{16}$  & ---  & 64  & (6,4,0) &  3,;414  & 1,;316 &    821,;248 & 157 & $57R$\\
 $2^{18}$  & ---  & 64  & (6,4,0) &  3,;548  & 1,;349 &  3,278,;848 & 140 & $51R$\\
 $2^{20}$  & ---  & 64  & (6,4,0) &  5,;956  & 1,;366 & 13,109,;248 & 102 & $38R$\\
\hline
 $2^{15}$  & ---  & 64  & (7,4,0) &  3,;927  & 1,;427 &    444,;672 & 153 & $57R$\\
 $2^{15}$  & ---  & 64  & (8,4,0) &  4,;141  & 1,;569 &    477,;696 & 154 & $57R$\\
 $2^{15}$  & ---  & 64  & (9,4,0) &  4,;673  & 1,;689 &    510,;720 & 159 & $59R$\\
\hline
 $2^{15}$  & ---  & 64  & (7,3,0) &  3,;725  & 1,;365 &    411,;904 & 153 & $57R$\\
 $2^{15}$  & ---  & 64  & (7,4,0) &  3,;927  & 1,;427 &    444,;672 & 153 & $57R$\\
 $2^{15}$  & ---  & 64  & (7,5,0) &  3,;731  & 1,;496 &    477,;440 & 155 & $57R$\\
\hline
\hline
 $2^{15}$  & ---  & 64  & (5,5,0) &  2,;811  & 1,;235 &    411,;392 & 169 & $63R$\\
 $2^{17}$  & ---  & 64  & (5,5,0) &  2,;714  & 1,;263 &  1,640,;192 & 160 & $58R$\\
\hline
\multicolumn{9}{|c|}{Decoder from \cite{Leroux_TSP_2012}}\\
\hline
$2^{15}$  & --- & 64 & (5,5,0) &  58,;480 &  33,;451 &   364,;288 &  66 & 31$R$\\
$2^{17}$  & --- & 64 & (5,5,0) & 221,;471 & 131,;764 & 1,445,;632 &  10 &  6$R$\\
\hline
\end{tabular}
}
\end{table}

\begin{table*}[t]
\centering
\caption{Schedule of the proposed semi-parallel architecture, with $N=8$ and $P=2$.}
\resizebox{18.2cm}{!}{
\renewcommand{\arraystretch}{1.1}
\begin{tabular}{|c||c|c|c|c|c|c|c|c|c|c|c|c|c|c|c|c|c||c|c|c|}
\hline
Stage / CC  & 0     & 1     & 2   & 3    & 4   & 5   & 6   & 7   & 8     & 9     & 10  & 11 & 12 & 13 & 14 & 15 & 16 & 17 & 18 & 19\\
\hline
$l=2$ & $f_0$ & $f_1$ &     &      &&     &      &&&& $g_0$ & $g_1$ &     &      &&     &      &&&\\
\hline
$l=1$ &       &       & $f$ &      && $g$ &      &&&&       &       & $f$ &      && $g$ &      &&&\\
\hline
$l=0$ &       &       &     & $fg$ &&     & $fg$ &&&&       &       &     & $fg$ &&     & $fg$ &&&\\
\hline
$l_\text{enc}=0$ &&&&& $e$ &&& $e$ &&&&&&&      $e$ &&& $e$ &&\\
\hline
$l_\text{enc}=1$ &&&&&&&&&        $e_0$ & $e_1$ &&&&&&& && $e_0$ & $e_1$\\
\hline
Output  &&&& $\Hu{0}\Hu{1}$ &&& $\Hu{2}\Hu{3}$ &&&&&&& $\Hu{4}\Hu{5}$ &&& $\Hu{6}\Hu{7}$ &&&\\
\hline
\end{tabular}
}
\label{tbl:schedule}
\end{table*}

\section{Experimental Results}
\label{sec:experimental_results}
The various characteristics of this architecture, explored in \secref{sec:architecture}, are summarized in \tblref{tbl:arch2_characteristics}. In this table, latency takes into account the semi-parallel schedule, as well as the chained PE and partial-sum encoder. The \emph{LLR SRAMs} entry combines both the channel and the internal LLR memories. $\Hs{}$ SRAMs store internal encoding estimations, but not the whole estimated codeword $\Hx{}$. Finally, throughput is estimated for $P=64$, a common value, to simplify its representation.

\tblref{tbl:fpga_results} then presents implementation results targeting an Altera Stratix~IV FPGA. Maximum frequencies are reported for the \emph{slow 900mV 85$^\circ$C} timing model. 

This table starts by presenting implementation results for the four $N=2^{15}$ codes described in \secref{sec:quantization}. It then explores the scalability of our design with respect to two parameters: code length and quantization. Finally, it compares this work with \cite{Leroux_TSP_2012}.

\begin{table}[t]
\renewcommand{\arraystretch}{1.5}
\centering
\caption{Summary of the technical characteristics of the proposed architecture.}
\resizebox{\columnwidth}{!}{
\begin{tabular}{|c||c|}
\hline
Decoding latency (CC) & $\frac{N}{P}(\frac{5P}{2}-1) + \frac{2N}{P} \log_2(\frac{N}{4P}) - \log_2 P + 2$\\
\hline
LLR SRAMs (bits)                                            & $Q_c N + Q(N + P\log_2 P - P)$\\
\hline
$\hat{s}$ SRAMs (bits)                                      & $P \big( \frac{3N}{2P} + 2\log_2 P - 4 \big)$\\
\hline
ROM (bits)                                                 & $N$\\
\hline
T/P [$P=64$] (bits/sec)      & $\sim R \frac{32}{71.5 + \log_2 N} f_\text{max} $ \\
\hline
\end{tabular}
}
\label{tbl:arch2_characteristics}
\end{table}

Note that, as in \cite{Leroux_TSP_2012}, our decoder architecture is not affected by code rate, as the choice of a specific code only modifies the contents of a ROM. Code rate is thus only reported in the first section of this table.

Those results show that the improved architecture retains a high clock frequency over a wide variety of code lengths, due to its fixed datapaths; the decreases observed are mostly due to routing delays, as more SRAM elements are used on the FPGA. Compared to \cite{Leroux_TSP_2012}, this new design scales much better with respect to all parameters; its higher memory use could be compensated, in an actual decoder, by $Q_{i_c} < Q_i$, while it is set to the same value here, for fair comparison.

The register, logic and memory use of the decoder targeting the $N=2^{20}$ code amount to 0.5\%, 2\%, and 72\% of the resources available on the selected FPGA, respectively. Additionally, register and logic use grow roughly linearly in the number of PEs and quantization bits, but are mostly unaffected by code length. Therefore, we can state that this architecture will scale to extremely long codes, limited almost exclusively by the amount of SRAM available on the FPGA.

At $N=2^{17}$, the largest code length supported by our previous-generation decoder, this improved architecture uses $81$ times less look-up tables (LUT), $104$ times fewer flip-flops (FF), has a maximum operating frequency $16$ times higher, and a throughput $11$ times greater, using the same parameters.

\section{Conclusion}
\label{sec:conclusion}
In this paper, we presented a scalable architecture for SC decoding of polar codes. This decoder features a semi-parallel, encoder-based partial-sum update module. This module utilizes SRAM for storage, and makes use of a fixed datapath. Additionally, this architecture leverages a multi-level quantization scheme for LLRs, decreasing memory use and decoder area. This state-of-the art decoder was synthesized for an Altera Stratix~IV FPGA target up to $N=2^{20}$, limited almost exclusively by the amount of available SRAM.

\section*{ACKNOWLEDGEMENT}
The authors would like to thank Gabi Sarkis and Pascal Giard, of McGill University, for helpful discussions.

\bibliographystyle{IEEEbib}
\bibliography{IEEEabrv,bibliography}

\end{document}

%% file: defines.tex
\newcommand{\Arikan}{Ar{\i}kan}

\newcommand{\citneed}[1]{{\color{red}\{Citation needed\}}\PackageWarning{CitationNeeded}{}}
\newcommand{\figref}[1]{Figure~\ref{#1}}
\newcommand{\secref}[1]{Section~\ref{#1}}
\newcommand{\tblref}[1]{Table~\ref{#1}}

\newcommand{\bfu}{\mathbf{u}}
\newcommand{\bfuhat}{\mathbf{\hat{u}}}
\newcommand{\bfx}{\mathbf{x}}
\newcommand{\bfxhat}{\mathbf{\hat{x}}}
\newcommand{\bfy}{\mathbf{y}}
\newcommand{\comment}[1]{}

\newcommand{\Hf}[1]{\hat{f}_{#1}}
\newcommand{\Hg}[1]{\hat{g}_{#1}}
\newcommand{\Hu}[1]{\hat{u}_{#1}}
\newcommand{\Hs}[1]{\hat{s}_{#1}}
\newcommand{\Hx}[1]{\hat{x}_{#1}}
\newcommand{\Mag}[1]{|#1|}

\newcommand{\Signr}[1]{\text{sign}(#1)}

\newcommand{\LLR}[1]{\lambda_{#1}}

\definecolor{deepgreen}{rgb}{0.1,0.6,0.1}
\definecolor{deeporange}{rgb}{0.8,0.4,0.1}
\definecolor{deeppurple}{rgb}{0.53,0.3,0.53}

%% file: fig-qtz_perf.tex
\begin{figure}%[!ht]
\centering
%\resizebox{2.68in}{!}{
\begin{tikzpicture}[every axis legend/.append style={nodes={right}}] 
\begin{semilogyaxis}[xlabel=$E_b/N_0$ (dB),ylabel=Error rate, grid=major, legend style={/tikz/every even column/.append style={column sep=0.25cm},at={(0.5,-0.2)},anchor=north}, legend columns=2, xtick={1,1.25,1.5,1.75,2},mark size=2.5pt, width=\columnwidth]

\addplot[color=blue,densely dashed,mark=triangle] coordinates {
    (1.00, 9.9220E-01)
    (1.25, 6.0010E-01)
    (1.50, 7.5200E-02)
    (1.75, 2.4168E-03)
    (2.00, 6.3705E-05)
}; \addlegendentry{FER/MSA, (6,3,2)}
\addplot[color=blue,mark=triangle] coordinates {
    (1.00, 4.7513E-01)
    (1.25, 2.4042E-01)
    (1.50, 2.4302E-02)
    (1.75, 6.1248E-04)
    (2.00, 1.1256E-05)
}; \addlegendentry{BER/MSA, (6,3,2)}
\addplot[color=red,densely dashed,mark=square] coordinates {
    (1.00, 9.8450E-01)
    (1.25, 5.1160E-01)
    (1.50, 4.8600E-02)
    (1.75, 2.0381E-03)
    (2.00, 2.3164E-05)
}; \addlegendentry{FER/MSA, float}
\addplot[color=red,mark=square] coordinates {
    (1.00, 4.7012E-01)
    (1.25, 2.0775E-01)
    (1.50, 1.6450E-02)
    (1.75, 6.1403E-04)
    (2.00, 4.4013E-06)
}; \addlegendentry{BER/MSA, float}
\addplot[color=black,densely dashed,mark=x] coordinates {
    (1.00, 9.1255E-01)
    (1.25, 2.9270E-01)
    (1.50, 1.9850E-02)
    (1.75, 4.4423E-04)
    (2.00, 1.0133E-05)
}; \addlegendentry{FER/SPA, float}
\addplot[color=black,mark=x] coordinates {
    (1.00, 4.0160E-01)
    (1.25, 1.0256E-01)
    (1.50, 5.5220E-03)
    (1.75, 1.0477E-04)
    (2.00, 1.1718E-06)
}; \addlegendentry{BER/SPA, float}

\end{semilogyaxis}
\end{tikzpicture}
%}
\caption{Performance of a $N=2^{15}$, $R=0.50$ polar code optimized for a frame error rate of $10^{-5}$.}
\label{fig:qtz_perf}
\end{figure}